\documentclass[preprint]{ptephy_v1}

\preprintnumber{KUNS-2748} 

\usepackage{graphicx}
\usepackage{epsf}
\usepackage{amsmath,amssymb}

\usepackage{float}
\usepackage{color} 
\usepackage{multirow}
\usepackage{dcolumn}
\usepackage{array}
\usepackage{comment}

\usepackage{ulem}

\def\vector#1{\mbox{\boldmath $#1$}}
\def\hana#1{\mathcal{#1}}
\def\Hesix{{}^6\textrm{He}}
\def\Beten{{}^{10}\textrm{Be}}
\def\2-b{\alpha + \Hesix}
\def\3-b{\alpha + \alpha + 2n}
\usepackage[utf8]{inputenc}

\usepackage{natbib}

\begin{document}
\title{Low-energy dipole excitation modes in $^{10}$Be}

\author{Yuki Shikata}
\author{Yoshiko Kanada-En'yo}
\author{Hiroyuki Morita}
\affil{Department of Physics, Kyoto University, Kyoto 606-8502, Japan}


\begin{abstract}
Low-energy dipole excitations in $\Beten$ were investigated 
based on the generator coordinate method with $\2-b$ and $\3-b$ cluster models.
We obtained three $1^-$ states in $E_{x}<15$~MeV, which show different characters in the 
$E1$, compressive dipole, and toroidal dipole transition strengths.
We found the strong toroidal dipole transition in the $1_1^-$ state, remarkable $E1$ strength in the
$1^-_2$ state, and the strong $E1$ and compressive dipole strength in the $1^-_3$ state.
The $1^-_1$ is described by the $\2-b$ cluster structure, where as the $1^-_2$ and $1^-_3$ states
are understood by three-body excitations of the $\3-b$ clustering. 
\end{abstract}

\subjectindex{D11, D13}

\maketitle

\section{Introduction}
In the dipole excitations of nuclei, significant low-energy dipole (LED) strengths have been observed in the 
energy region below the giant dipole resonance (GDR).
The LED strengths have been attracting a great interest in the experimental~\cite{1402-4896-2013-T152-014012,Bracco:2015hca} and theoretical sides~\cite{Paar:2007bk}, 
but their properties and origins have not been clarified yet. 

In studies of neutron-rich nuclei in these two decades, the isovector dipole (IVD), the so-called $E1$ excitations, 
in the low-energy region have been intensively investigated, and often discussed in astrophysical interests, for instance, 
in relation with photo-absorption in the $r$-process~\cite{Goriely:1998utv,Goriely:2004qb}.
Not only in the IVD excitations, but also in the isoscalor dipole (ISD) channel, significant low-energy strengths 
have been experimentally known in $Z=N$ stable nuclei 
as observed in ${}^{12}$C~\cite{John:2003ke}, ${}^{16}$O~\cite{Harakeh:1981zz}, ${}^{24}$Mg~\cite{Youngblood:1999zz}, ${}^{28}$Si~\cite{Youngblood:2002mk}, and ${}^{40}$Ca~\cite{Poelhekken:1992gvp}. 
A recent focus is their isospin property, 
which can be a key feature to understand the phenomena of the LED excitations. 

The significant LED strengths may indicate emergence of possible new excitation modes 
different from the GDR modes, which are usually understood by the collective oscillations of 
the whole system such as the proton-neutron incoherent oscillation for the IV-GDR and the compressive dipole oscillation for the IS-GDR. 
A couple of dipole excitation modes have been suggested to describe the LED strengths. 
For the LED in neutron-rich nuclei, the so-called ``pigmy mode'' has been considered to describe the low-energy $E1$ strengths~\cite{Ikeda:pygmy}. 
It is characterized by weakly bound valence neutron motion in neutron skin or neutron-halo structure against a core.
Indeed, the pigmy mode has been theoretically predicted by many theoretical studies with a fluid dynamical model~\cite{Mohan:1971tz,VanIsacker:1992zz,10.1143/PTP.83.180}
and microscopically mean-field approaches\cite{Vretenar:2001hs,Colo:2001fz,Sarchi:2004pf,Cao:2005bt,Terasaki:2006ts,Yoshida:2008rw,Inakura:2011mv,Inakura:2018ccl}.

Another candidate for the LED is the ``toroidal mode", which is also called as the ``torus mode'' or 
``vortical mode'' \cite{Dubovik:toroidal,Dubovik:toroidal2}.
The toroidal dipole mode is characterized by the vorticity of nuclear current and is the counter part of the 
standard compressive dipole mode for the IS-GDR as originally proposed by nuclear fluid-dynamics
to understand the IS-LED strengths in the stable nuclei ~\cite{Semenko:toroidal,Bastrukov:1992ep}. 
In this decade, the vortical nature of the toroidal mode have been microscopically studied with mean-field approaches
mainly for spherical nuclei in the heavy-mass region \cite{Vretenar:2001te,Ryezayeva:2002zz,0954-3899-29-4-312,Kvasil:2011yk,Repko:2012rj,Reinhard:2013xqa}. 
Recently, theoretical studies of the toroidal mode have been extended to deformed nuclei, in which 
further rich phenomena are expected because of coupling of the LED mode with the nuclear 
deformation~\cite{Kvasil:2013soa,Kvasil:2013yca,Nesterenko:2016qiw,Repko:2017uvc,Nesterenko:2017rcc}.
In our recent works with the antisymmetrized molecular dynamics (AMD),  
we have studied the LED in light nuclei such as ${}^{10}$Be~\cite{Kanada-Enyo:2017uzz} 
and ${}^{12}$C~\cite{Kanada-Enyo:2017fps}, and  
predicted the toroidal dipole excitations.
We have also shown that the cluster structure plays an important role in the LED of light nuclei. 

In general, cluster structures can contribute to the low-energy ISD excitations in light nuclei,
 because the ISD operator, which is the compressive type with the $r^3$ term,  
can directly excite the inter-cluster motion as pointed out by Chiba {\it et al.} in a similar way to the IS monopole excitations with the $r^2$ term~\cite{Kawabata:2006wy,Yamada:2007ua,Chiba:2015khu,Chiba:2016zyz}.
It means that the cluster excitation mode can be another candidate for the LED excitations.

In this paper, we investigate the LED excitation modes in ${}^{10}$Be in terms of cluster dynamics using generator coordinate method (GCM) with cluster model wave functions.
In our previous work~\cite{Kanada-Enyo:2017uzz}, we showed that the $1_1^-$ and $1_2^-$ states have remarkable LED strengths based on antisymmetrized molecular dynamics(AMD) combined with the generator coordinate method (GCM).
However, the previous model includes only the $\alpha$-$\Hesix$ relative motion but does not treat the internal excitation of $\Hesix$ cluster.
In the present calculation, we use $\2-b$ and $\3-b$ cluster model functions and discuss not only 2-body but 3-body cluster dynamics.
As a result, we obtain three $1^-$ states having remarkable LED strengths in $E<15$ MeV.
The $1_1^-$ has the toroidal nature which is consistent with the $1_1^-$ state of the previous result.
The $1_2^-$ and $1_3^-$ states have significant $E1$ strength and the $1_3^-$ has significant CD strength.
We discuss the contribution of $\alpha-$ and $2n$-cluster modes to the properties of these states.
The mechanism of the toroidal mode in the plolately deformed system is also discussed.

This paper is organized as follows.
The framework of the present cluster model with the GCM is explained in the next section, 
and the calculated results are shown in Sec.~\ref{sec:result}. 
Sec.~ \ref{sec:discussion} discusses cluster features of the LED. Finally, a summary is given 
in Sec.~\ref{sec:summary}.

\section{Formulation}\label{sec:formulation}
In the present study of dipole excitations in  $\Beten$, we  apply a cluster model with the GCM to 
calculate the ground and $1^-$ excited states of $\Beten$.
For the basis cluster wave functions, two kinds of wave functions 
are adopted and superposed in the GCM calculation. 
One is the di-cluster wave function of $\alpha+{}^6$He  which 
has been used in Ref.~\cite{Kanada-Enyo:2011ldi}. 
The other is the tri-cluster wave function of $\3-b$. 
The former is suitable to describe the two-body excitation mode of the $\alpha+{}^6$He, and the latter takes into account three-body dynamics of the $\alpha+\alpha+2n$ in the dipole excitations. 
In this section, we explain details of the model wave functions and procedures of the present calculation.

\subsection{basis wave functions}\label{sec:cluster_model}
The cluster wave functions are given by the Brink-Bloch (BB) cluster wave functions~\cite{bloch1966many}.
The BB cluster wave function for a system consisting of $C_1,\ldots,C_m$ clusters ($m$ is the number of 
constituent clusters) is given as 
\begin{eqnarray}
\Phi_{C_1+\cdots+C_m}(\vector{S}_1,\ldots,\vector{S}_m) 
= \hana{A}\left[ \Phi_{C_1}(\vector{S}_1)\cdots\Phi_{C_m}(\vector{S}_m) \right],
\end{eqnarray}
where $\hana{A}$ is the antisymmetrizer, and $\Phi_{C_i}$ is the wave function of the $C_i$ cluster placed at the mean position $\vector{S}_i$. 
For cluster positions, the condition of $\sum_i A_i\vector{S}_i /A=0$ is fulfilled so as to exactly remove the total center of mass motion. 
Here $A_i$ is the mass number of the $C_i$ cluster.  

\begin{figure}[!h]
\begin{center}
\includegraphics[width=10cm]{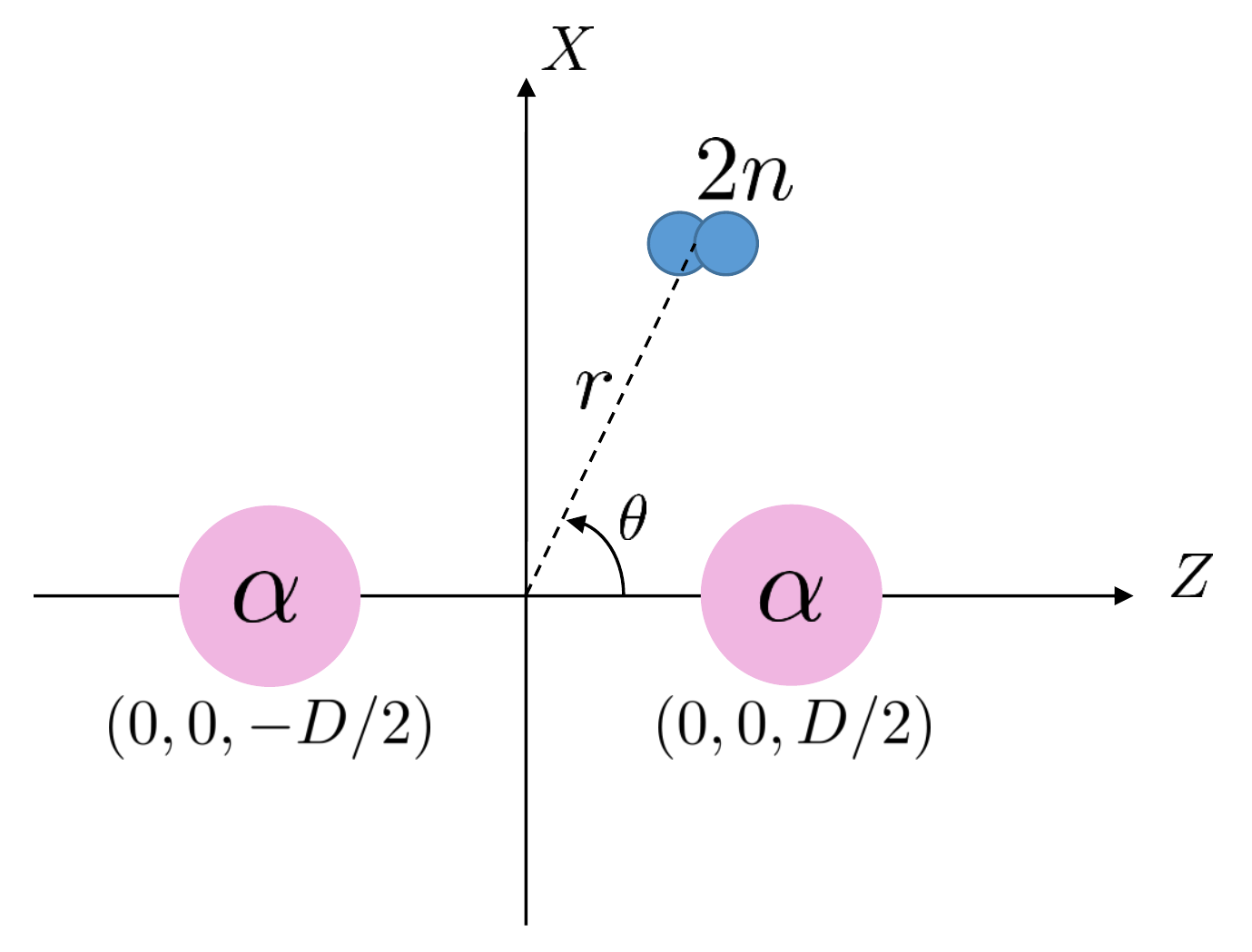}
\end{center}
\caption{(color online) Schematic figure for definitions of the parameters $D$, $r$, and $\theta$ in the $\3-b$ model. 
}
\label{fig:3-body_parameter}
\end{figure}

For the tri-cluster wave function of $\3-b$, we assume the $\alpha$- and $2n$-cluster wave functions to be 
the harmonic oscillator $(0s)^4$ and $(0s)^2$ configurations, respectively, and take a common oscillator width. 
The BB wave function for the $\3-b$ cluster  is given as 
\begin{eqnarray}
\Phi_{\3-b}(D,r,\theta) &=& \hana{A}\left[ \Phi_{\alpha}(\vector{S}_1)\Phi_{\alpha}(\vector{S}_2)\Phi_{2n}(\vector{S}_3) \right], \\
\Phi_{2n}(\vector{S}) &=& \hana{A}[\psi_{n\uparrow} (\vector{S})\psi_{n\downarrow}(\vector{S})].
\end{eqnarray}
The cluster position parameters are set as $\vector{S}_1=(0,0,-D/2)$, $\vector{S}_2=(0,0,D/2)$, 
and $\vector{S}_3=(r\sin\theta,0,r\cos\theta)$, where $(D,r,\theta)$ are introduced as the model parameters of the generator coordinates 
in the $\3-b$ cluster model. Here, 
$D$ is the distance between the two $\alpha$-clusters,
$r$ is the distance between the $2n$-cluster and the mass center of $2\alpha$, and
$\theta$ indicates the direction of the vector $\vector{r}$ relative to the $\alpha -\alpha$ axis.
A schematic figure of these parameter settings is shown in Fig.~\ref{fig:3-body_parameter}.

For the $\2-b$ cluster wave function, the ${\Hesix}$ cluster is given by the harmonic oscillator $p^2$ configurations with the 
oscillator width same as that of the $\alpha$-cluster. We adopt the same $\2-b$ cluster wave function used in 
Ref.~\cite{Kanada-Enyo:2011ldi} as 
\begin{eqnarray}
\Phi_{\alpha + {}^6\textrm{He}}(D, \sigma) = \hana{A}[\Phi_{\alpha}(\vector{S}_1)\Phi_{\Hesix}(\vector{S}_2, \sigma)], 
\end{eqnarray} 
where $D$ is the relative distance between two clusters, and 
$\sigma$ represents the $p$-shell configuration of two valence neutrons of ${\Hesix}$. 
Practically, six configurations $\sigma=1,\ldots,6$ are superposed in order to describe all $(0p)^2$ configurations 
of valence neutrons coupled to be $0^+$ and $2^+$ states of ${\Hesix}$ 
in the $\alpha + {}^6\textrm{He}$ wave function (see  Ref.~\cite{Kanada-Enyo:2011ldi} for the details of configurations).
It should be commented that the $\alpha + {}^6\textrm{He}$ model treats each configuration of $(p_{3/2})^2$ and 
$(p_{1/2})^2$ as well as their mixing in the  ${}^6\textrm{He}$ cluster. 

\subsection{Superposition with GCM}\label{sec:GCM}
In order to obtain the total wave function of the ground and $1^-_k$ states of $\Beten$, we superpose the parity and 
angular-momentum projected tri-cluster ($\Phi_{\3-b}$) and di-cluster ($\Phi_{\2-b}$)
wave functions with the GCM with respect to the parameters and configurations 
as  
\begin{eqnarray}
\Psi_{M}^{J_k\pi} &= \sum_{D,r, \theta}\sum_K c^{(\textrm{tri})}_{D,r,\theta;K}
\hat{P}_{MK}^{J}\hat{P}^{\pi}\Phi_{\3-b}(D,r,\theta)\nonumber\\
&+\sum_{D,\sigma}\sum_K c^{(\textrm{di})}_{D;\sigma, K}\hat{P}_{MK}^{J}\hat{P}^{\pi}\Phi_{\alpha + {}^6\textrm{He}}(D, \sigma),  \label{eq:total_wf}
\end{eqnarray}
where $\hat{P}_{MK}^{J}$ and $\hat{P}^{\pi}$ are the angular-momentum and parity projection operators.
The generator coordinates $D$, $r$, and $\theta$ are discretized, and 
the coefficients $c^{(\textrm{tri})}_{D,r,\theta;K}$ and $c^{(\textrm{di})}_{D;\sigma,K}$ are determined by 
diagonalization of Hamiltonian and norm matrices. 

As a base of the GCM, these parameters are taken as  $D = 1,2,\cdots,8,$ and $\sigma=(\textrm{6 configurations})$ in the $\2-b$ model and $D = 2,3,\cdots ,6$, $r = 0.4, 1.4,\cdots , 5.4$ and $\theta = 0,\frac{\pi}{8},\cdots ,\frac{\pi}{2}$ in the $\3-b$ model.
Therefore, totally $6\times 8 + 5\times 6\times 5 = 198$ bases are taken.
As a width parameter we set $\nu=0.235\ \textrm{fm}^{-2}$ as same as that used in Ref.~\cite{Kanada-Enyo:2011ldi}.

\subsection{dipole transition strengths}\label{sec:dipole_op}
In order to investigate properties of the LED states, we calculate the dipole transition strengths from the ground state to the $1_{k}^-$ states obtained with the GCM calculation.
The $E1$ operator for the IVD transitions is given as  
\begin{eqnarray}
\hat{M}_{E1}(\mu) =  \frac{N}{A}\sum_{i\in p}r_iY_{1\mu}(\hat{\vector{r}}_{i}) - \frac{Z}{A}\sum_{i\in n}r_iY_{1\mu}(\hat{\vector{r}}_{i}).
\end{eqnarray}
For the IS dipole transitions, the compressive dipole (CD) operator is given as 
\begin{eqnarray}
\hat{M}_{\textrm{CD}}(\mu) &=& \frac{-1}{10\sqrt{2}c}\int d\vector{r}\ \nabla\cdot\vector{j}_{\textrm{nucl}}(\vector{r})\ r^3Y_{1\mu}(\hat{\vector{r}}), \\
\vector{j}_{\textrm{nucl}}(\vector{r}) &=& \frac{-i\hbar}{2m}\sum_{k=1}^A\{ \vector{\nabla}_k\delta(\vector{r}-\vector{r}_k) + \delta(\vector{r}-\vector{r}_k)\vector{\nabla}_k \}, \label{eq:current}
\end{eqnarray}
where $\vector{j}_{\textrm{nucl}}(\vector{r})$ is the convection nuclear current.
The CD operator is the higher order $r^3$  term because the lowest $r$ term is just the 
translational operator of the center of mass motion and does not contribute to the ISD transitions in nuclei. 
As shown later, the CD transition strength corresponds to the ISD transition strength with the ordinary ISD operator 
\begin{eqnarray}
\hat{M}_{\textrm{ISD}}(\mu)=\int d\vector{r}\ \rho(\vector{r})r^3Y_{1\mu}(\hat{\vector{r}}), 
\end{eqnarray}
and is a good probe for the compressional dipole mode. 
In addition to the CD operator, the following toroidal dipole (TD) operator is considered,
\begin{eqnarray}
\hat{M}_{\textrm{TD}}(\mu)=\frac{-1}{10\sqrt{2}c} \int d\vector{r}\ (\hat{\nabla}\times\vector{j}_{\textrm{nucl}}(\vector{r}))\cdot r^3\vector{Y}_{11\mu}(\hat{\vector{r}}),
\end{eqnarray}
where $\vector{Y}_{jL\mu}(\hat{\vector{r}})$ is vector spherical as follows,
\begin{eqnarray}
\vector{Y}_{jL\mu}(\hat{\vector{r}}) = \sum_{\alpha,\beta}\langle L\alpha,1\beta|j\mu\rangle Y_{L\alpha}(\hat{\vector{r}})\vector{e}_\beta ,
\end{eqnarray}
where $\vector{e}_\beta$ is unit vector in spherical basis.
The TD operator can sensitively probe nuclear vorticity and is a counter part of the 
CD operator~\cite{0954-3899-29-4-312}.

For these three types of the dipole operators, $\hat{M}_D=\{\hat{M}_{E1}, \hat{M}_\textrm{CD}$, $\hat{M}_\textrm{TD}$\},
transition strength of the dipole operator $\hat{M}_D$ from the ground state
is given as
\begin{eqnarray}
B(D;0_1^+\rightarrow 1_k^-) = |\langle 1_k^-||\hat{M}_D||0_1^+\rangle|^2,
\end{eqnarray}
where $\langle 1_k^-||\hat{M}_D||0_1^+\rangle$ is the reduced matrix element. 
The CD transition strength is related to the ordinary ISD transition strengths as 
\begin{eqnarray}
B(\textrm{CD};0_1^+\rightarrow 1_k^-) &=& \left(\frac{1}{10}\frac{E_k}{\hbar c}\right)^2B(\textrm{ISD};0_1^+\rightarrow 1_k^-), 
\end{eqnarray}
where $E_k$ is the excitation energy for the $1_k^-$ state and 
$B(\textrm{ISD};0_1^+\rightarrow 1_k^-) = |\langle 1_k^-||\hat{M}_\textrm{ISD}||0_1^+\rangle|^2$.

\section{Result}\label{sec:result}
We apply the GCM with the $\alpha + {}^6$He and $\alpha + \alpha + 2n$ cluster wave functions, 
to investigate the LED excitation modes in ${}^{10}$Be. 
We show the dipole transition strengths and discuss the properties of 
dipole excitations in $E < 15$ MeV while focusing on the cluster structures.

\subsection{Effective interactions}
In the present study, effective two-body nuclear forces including the central ($v^\textrm{central}_{ij}$) and 
spin-orbit ($v^{ls}_{ij}$) interactions are used.
Hamiltonian of the total system is given as 
\begin{eqnarray}
H = \sum_i T_i - T_G + \sum_{i<  j}(v^\textrm{central}_{ij} + v^{ls}_{ij} + v^C_{ij}), 
\end{eqnarray} 
where $T_i$ and $T_G$ are the kinetic energy of the $i$th nucleon and that of the total center of mass motion, respectively.
$v^C_{ij}$ is the Coulomb interaction, which is approximated by a seven-range Gaussian function.
As for parameters of the nuclear interactions, we use the same parametrization as Refs.~\cite{Suhara:2009jb,Kanada-Enyo:2011ldi}: 
the Volkov No.2 force\cite{Volkov:1965zz} with $W = 1-M=0.6$ and  $B=H=0.125$ for 
the central interaction ($v^\textrm{central}_{ij}$) and the G3RS force\cite{Yamaguchi:1979hf,Tamagaki:1968zz} with the strengths
$u_1 =-u_2 = -1600$ MeV for the spin-orbit interaction ($v^{ls}_{ij}$). 
These parameters reproduce the properties of 
sub systems such as the $\alpha - \alpha$ and $\alpha - n$ scattering phase shifts as well as $S$-wave nucleon-nucleon 
scattering phase shifts.

\subsection{LED states and transition strengths}\label{sec:strengths}
\begin{figure}[!h]
\begin{center}
\includegraphics[width=10cm]{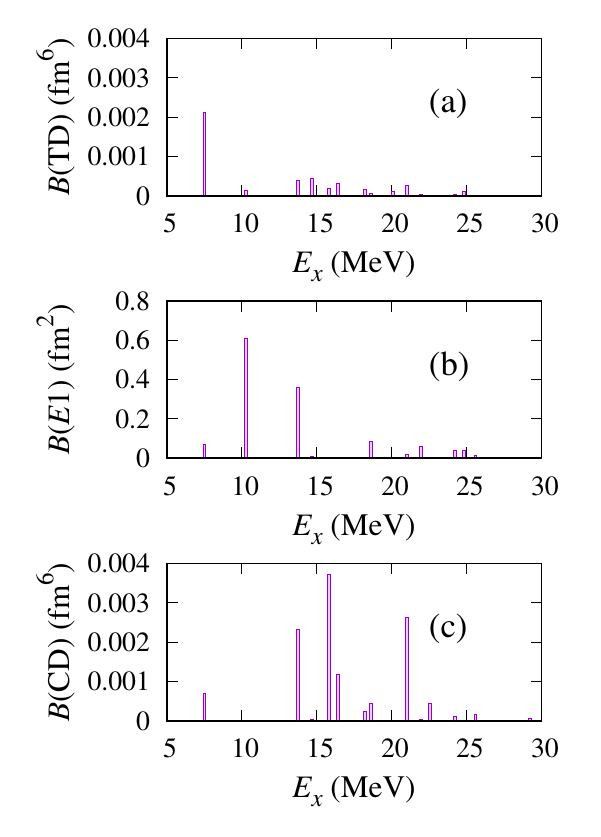} 	
\end{center}
\caption{(color online) Dipole transition strengths for the (a) TD, (b) $E1$, and (c) CD operators 
for $1^-_k$ states up to $30$ MeV calculated with the full GCM calculation. 
}
\label{fig:full_strength}
\end{figure}

With the GCM calculation, we obtain the binding energy $58.48$~MeV
of $\Beten$, which more or less underestimates the experimental value $65.0$~MeV.
Figure \ref{fig:full_strength} shows the energy spectra and the transition strengths of dipole excitations up to $30$ MeV.
In the TD, $E1$, and CD transition strengths in the low-energy $E_x < 15$~MeV region, we obtain 
three $1^-$ states (the $1^-_1$, $1^-_2$, and $1^-_3$ states) 
having strong dipole transitions,  which we call the LED states.
In the $E_x > 15$~MeV region, there are no significant strengths of the TD and $E1$ transitions. 
The present model is based on the cluster model and is not enough to describe the IVGDR because
internal excitations of clusters are omitted. 
In the CD transitions, we obtain some strengths. 
These high-energy CD strengths are considered to be partial contributions to the ISGDR from 
the inter-cluster motion.

The lowest state, $1_1^-$, obtained at $7.52$~MeV is assigned to the observed $1_1^-$ at $5.96$~MeV~\cite{ANDERSON197477,AJZENBERGSELOVE19741}.
This state has the remarkably large TD strength and therefore we call this state the TD state.
This TD state also has the finite CD strength as $3.4 \%$ of the energy weighted sum rule (EWSR), but the small $E1$ strength.
The weak $E1$ transition of the  $1_1^-$ is qualitatively consistent with the observation, but quantitatively, 
the calculated value $EB(E1;0_1^+\rightarrow 1_1^-) = 5.38 \times 10^{-1}$~fm${}^2$MeV overestimates 
the extremely small experimental value, $EB(E1;0_1^+\rightarrow 1_1^-) = 1.62\times 10^{-5}~\textrm{fm}^2\textrm{MeV}$~\cite{PhysRevC.80.034318}.

In $E_x=10 - 15$~MeV region, we obtain two $1^-$ states with remarkable $E1$ strengths: 
the $1_2^-$ at $10.27$~MeV with $17.5~\%$ and the $1_3^-$ at $13.78$~MeV with $13.9~\%$ of the 
Thomas-Reich-Kuhn (TRK) sum rule $S(\textrm{TRK}) = \frac{9\hbar^2}{8\pi M}\frac{NZ}{A}\sim35.64~\textrm{fm}^2\textrm{MeV}$.
In the CD transition strengths, one can see a difference between these two LED states. 
The remarkably strong CD strength is obtained for the $1_3^-$ as $6.2~\%$ of the EWSR, but the CD strength almost vanishes in 
the $1_2^-$.
In the following, we call the $1_2^-$ and $1_3^-$ states $E1$ and CD states, respectively. 

Owing to the contributions from the three LED states,  
the energy weighted sum (EWS) in $E_{x}<15$~MeV region is  $\sim~10 \%$ of the EWSR for the CD transitions and 
 $33~\%$ of the TRK sum rule for the $E1$ transitions.


\section{Discussion}\label{sec:discussion}

\subsection{Contributions of $\2-b$ and $\3-b$ configurations}\label{sec:cluster_strength}

\begin{figure}[!h]
\begin{center}
\includegraphics[width=15cm]{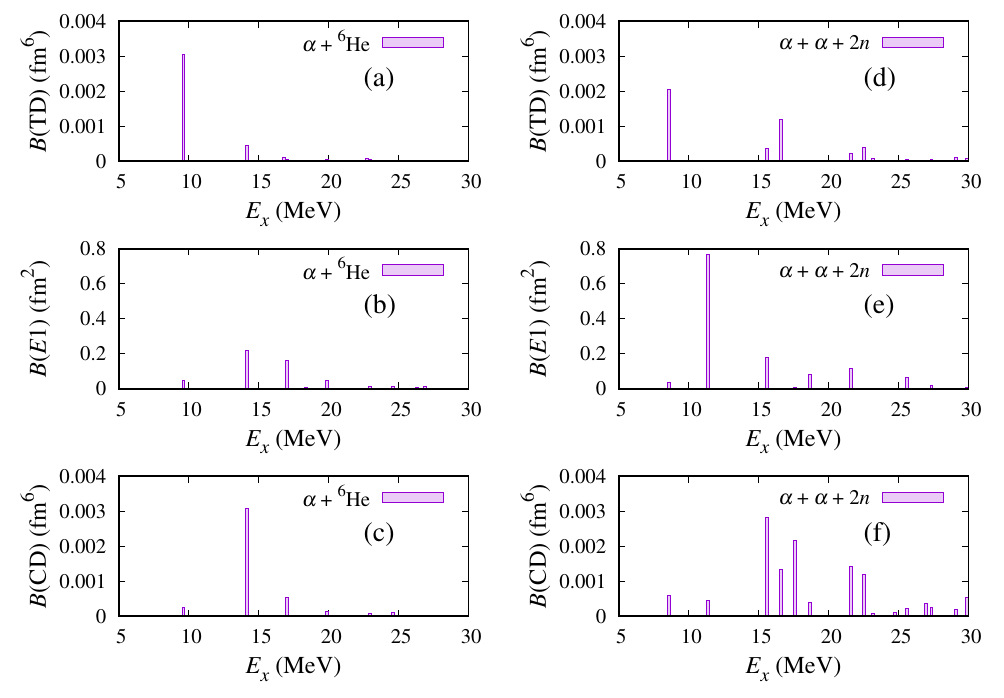} 	
\end{center}
\caption{(color online) Dipole transition strengths for the TD, $E1$, and CD operators obtained by the 
(a)-(c) $\2-b$ and (d)-(e) $\3-b$ calculations.
}
\label{fig:two_three_body}
\end{figure}

In order to investigate cluster structures of the LED states and their roles in the dipole transition strengths, 
we make further analysis by truncating the model space of the GCM bases.
We perform the GCM calculation only with the di-cluster ($\2-b$) configurations and that only with the tri-cluster ($\3-b$) 
configurations to separately see contributions from 2-body and 3-body cluster modes. 
The results are shown in Fig.~\ref{fig:two_three_body}. 
Figures~\ref{fig:two_three_body} (a)-(c) show the TD, $E1$, and CD transition strengths obtained by the $\2-b$ calculation,
and Figures~\ref{fig:two_three_body} (d)-(f) show the strengths obtained by the $\3-b$ calculation.
In the following discussions, we label $1^-_k$ states obtained with the  $\2-b$ calculation as $1^-_{k,\textrm{di}}$,
and those obtained with the $\3-b$ calculation as $1^-_{k,\textrm{tri}}$.
The excitation energies and transition strengths of the $1^-_{k,\textrm{di}}$ and $1^-_{k,\textrm{tri}}$ states are
calculated for the $0_1^+$ obtained by the full GCM calculation.

In the $\2-b$ calculation, we obtain the $1_{1,\textrm{di}}^-$ state with the strong TD strength at $E_{x}\sim 10$ MeV, which 
corresponds to the TD state ($1_{1}^-$) of the full GCM calculation. 
It means that the TD state is dominantly described by the $\2-b$ configurations.
The TD state is obtained also in the $\3-b$ calculation because the $\2-b$ configurations are partially included in the $\3-b$ model space. 
Comparing the results between the $\2-b$ calculation in Figs.~\ref{fig:two_three_body}(a)-(c)
and the GCM calculation in Fig.~\ref{fig:full_strength},
one can see that the excitation energy of the TD state is lowered by 2~MeV and that the CD strengths of the TD state is slightly increased by a factor of 3
in the full GCM calculation because of inclusion of $\3-b$ configurations into $\2-b$ configurations.

For the $E1$ and CD states ($1_2^-$ and $1_3^-$), corresponding states are not obtained in 
the $\2-b$ calculation. There is no state with strong $E1$ transition in the low-energy region.
At $E_{x}\sim 15$~MeV,  a $1^-$ state with the significant CD strength 
is obtained as the $1^-_{2,\textrm{di}}$ but it does not correspond to the CD state ($1_3^-$)
because it has small overlap.
On the other hand, in the $\3-b$ calculation, we obtain the two $1^-$ states in $10\lesssim E_{x} \lesssim 15$~MeV region as
the $1^-_{2,\textrm{tri}}$ and  $1^-_{3,\textrm{tri}}$ states, which contribute to the dominant components of the  $E1$ and CD states. 
The $1^-_{2,\textrm{tri}}$ at $E_{x}\sim 11$ MeV and the $1^-_{3,\textrm{tri}}$ at  $E_{x}\sim 15$~MeV 
show remarkable $E1$ and CD transition strengths, respectively, and have significant overlap with 
the $E1$ and CD states ($1_2^-$ and $1_3^-$) obtained in the full GCM calculation. 
This result indicates that dominant components of the $E1$ and CD states are contributed by the tri-cluster $\3-b$ mode.
However the coupling with the $\2-b$ configurations affects detailed properties of the $E1$ and CD states.
In the $\3-b$ calculation, 
the $E1$ transition strength is concentrated on the $1_{2,\textrm{tri}}^-$, 
but the strength in the full GCM calculation is fragmented into the $1^-_2$ and $1^-_3$ states because of the mixing of the $\2-b$ configurations.

In the following sections, we discuss spatial development of the cluster structures in the LED states.
Particular attentions are paid on the $\alpha$-cluster  development evaluated by the $\alpha - \alpha$ distance 
and also on the $2n$-cluster development characterized by the spatial 
extent of the $2n$-cluster distribution from the $2\alpha$.
The former ($\alpha$-cluster development) is taken into account in both of the
$\2-b$ and $\3-b$ configurations, and the latter ($2n$-cluster development) is 
treated only with the $\3-b$ configurations

\subsection{$\alpha$-cluster developments in the LED states}\label{cluster_mode}

\begin{figure}[!h]
\begin{center}
\includegraphics[width=\hsize]{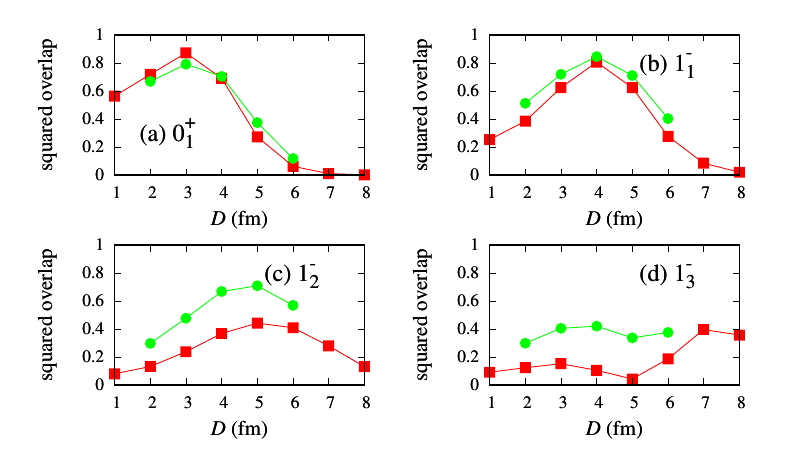}
\end{center}
\caption{(color online) The $\alpha$-cluster distributions of the $\2-b$ and $\3-b$ components in the 
(a)$0_1^+$, (b)$1_1^-$,  (c)$1_2^-$, and (d) $1_3^-$ states obtained by the full GCM calculation. 
The $\2-b$ and $\3-b$ components are plotted as functions of the $\alpha - \alpha$ distance $(D)$ by 
circle (red) and square (green) points.}
\label{fig:D_subspace}
\end{figure}

In order to see the $\alpha$-cluster development in the ground and LED states obtained by the GCM calculation, 
we calculate the squared overlap with the subspace of $\2-b$ configurations with a given value of the distance $D$ 
and that of $\3-b$ configurations. 
$D$ indicates the $\alpha - ^6$He distance of the $\2-b$ configurations and the 
$\alpha - \alpha$ distance of the  $\3-b$ configurations. The squared overlap
shown in Fig.~\ref{fig:D_subspace} indicates how much the 
$\2-b$ (square points) and $\3-b$ (circle points) components with a given distance $D$ are contained in the $0_1^+$, $1_1^-$, $1_2^-$, and $1_3^-$ states.

The ground state ($0_1^+$) shows the $\2-b$ nature with the relatively weak $\alpha$-cluster development as shown in the 
dominant $\2-b$ component with the maximum peak at $D = 3$ fm (see Fig.~\ref{fig:D_subspace} (a)).
The $\3-b$ component also shows 
the similar $D$ dependence just because the $\3-b$ model space includes a part of $\2-b$ model space. 

In the squared overlap for the LED states, 
one can see the larger $\alpha$-cluster development
compared with the $0_1^+$. 
The $1_1^-$ for the TD state has the maximum peak at $D=4$ fm (see Fig.~\ref{fig:D_subspace} (b)) 
slightly larger position than the $0_1^+$ case. 
The TD state has the dominant $\2-b$ component consistently with the analysis in the previous section.
In contrast to the dominant $\2-b$ component in the $1_1^-$ state, 
the $1_2^-$ and $1^-_3$ states have significant $\3-b$ component and minor $\2-b$ component. 
The $1_2^-$ for the $E1$ state shows the 
further developed $\alpha$-cluster with the 
maximum peak at  $D=5$ fm (see Fig.~\ref{fig:D_subspace} (c)) 
indicating that the dipole excitation to the $1_2^-$ is $\alpha$-cluster excitation.
In the $1_3^-$ for the CD state, significant amplitudes of the $\3-b$ component 
are distributed in a wide range of $D$ (see Fig.~\ref{fig:D_subspace} (d)). 

\subsection{Contribution of large amplitude $\alpha$-cluster motion}
\begin{figure}[!h]
\begin{center}
\includegraphics[width=15cm]{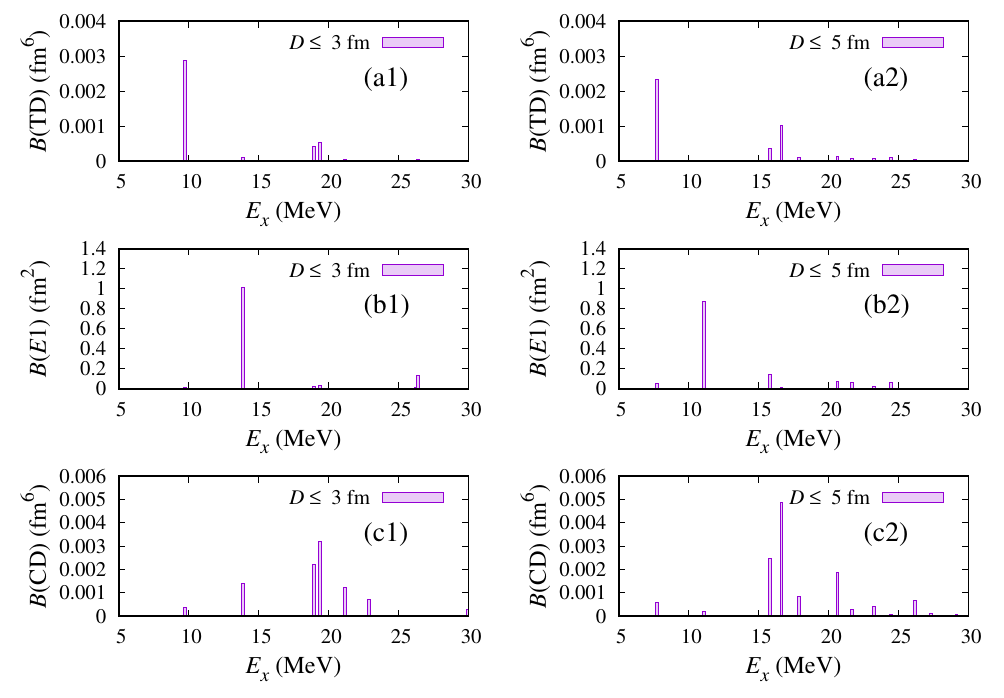} 	
\end{center}
\caption{(color online) TD, $E1$, and CD transition strengths obtained by the $D\le 3$~fm and $D\le 5$~fm calculations. }
\label{fig:D_truncated}
\end{figure}

We here discuss contribution of the large amplitude cluster motion to the LED states
based on the calculation with truncation of the $D$ ($\alpha - \alpha$ distance) space. 
We perform the GCM calculation using the $\2-b$ and $\3-b$ basis configurations with $D\le 3$~fm and that with $D\le 5$~fm and compare the results with the full GCM calculation. 
The obtained TD, $E1$, and CD strength functions are shown in Fig.~\ref{fig:D_truncated}.

For the TD state ($1_1^-$), it is found that the $D=4 - 5$~fm configurations contribute to lowering the excitation energy about 3.5~MeV
as seen in the $D\le 5$~fm case, whereas  $D>5$~fm configurations give almost no contributions. 
The  $D=4 - 5$~fm configurations also affect to reduce the TD transition strengths by $\sim~30~\%$ because 
the TD transition is sensitive to the surface nuclear current and weakens as the overlap with the $0^+_1$ state decreases in the $\alpha$-cluster developing.

For the $E1$ state ($1_2^-$), large $D$ configurations in both $D=4 - 5$~fm and $D>5$~fm regions 
play an important role in lowering the excitation energy. 
They contribute about $4$~MeV energy gain of the $1_2^-$ 
as seen in the comparison with the full GCM result.

As for the CD state ($1_3^-$) obtained at $E_{x}=13.5$~MeV by the full GCM calculation, it is difficult to make clear assignment, 
but some states in higher energy regions of the $D\le 3$~fm and $D\le 5$~fm calculations has significant overlap
with the CD state: a couple of
states around $E_{x}=20$~MeV of the $D\le 3$~fm calculation ($E_{x}=15 - 17$~MeV of the $D\le 5$~fm calculation).
It means that large $D$ configurations are essential to generate the LED state that has the strong CD transition.

We should comment another role of the large $D$ configuration to the CD strengths in the TD state ($1_1^-$).
As shown in comparison between the $D\le 3$~fm and $D\le 5$~fm calculations, the CD transition strength
of the TD state is increased by a factor of two by the $D = 4 - 5$ fm configurations. 
Even though the resultant CD strength is not so strong, the slight increase of the CD strength 
can be an indirect signature of the cluster structure of the TD state in the current situation,
 where the direct measurement of the TD state in neutron-rich nuclei is not feasible yet.
 
\subsection{$2n$-cluster development}
\begin{figure}[!h]
\begin{center}
\includegraphics[width=10cm]{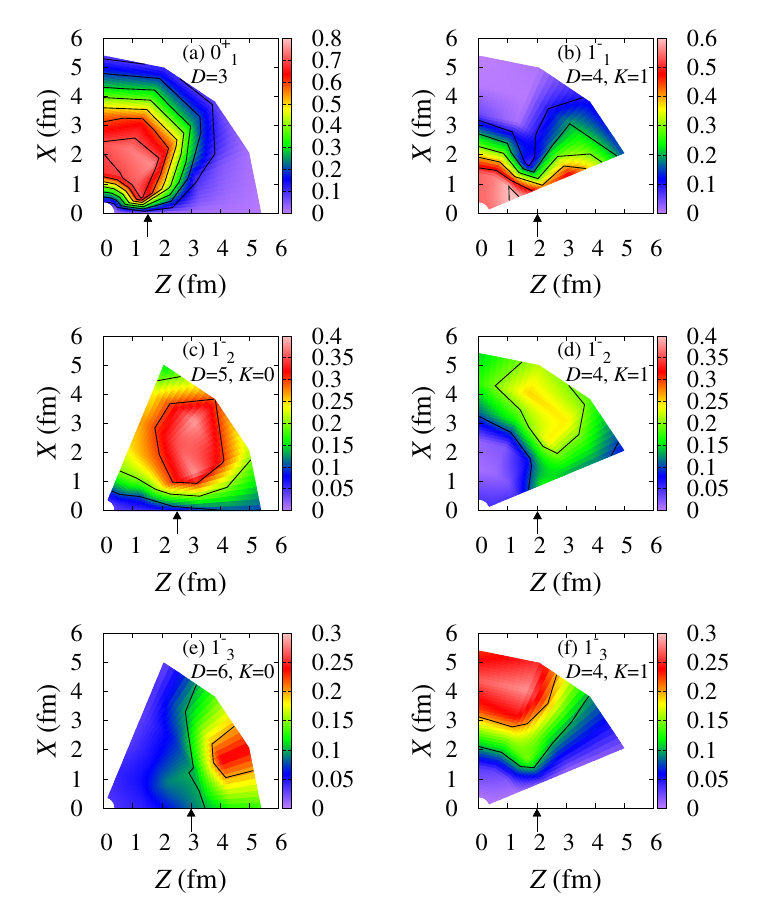} 	
\end{center}
\caption{(color online) The $2n$-cluster distributions for the $0_1^+,\ 1_1^-,\ 1_2^-,$ and $1_3^-$ states in the $\3-b$ configuration.
These distributions are calculated at given $D$ and $K$.
The panels (a) and (b) show the distribution for the $0_1^+$  at $(D=3~\textrm{fm},K=0)$ and for the $1_1^-$ at $(D=4~\textrm{fm},K=1)$.
For the $1_2^-$($1_3^-$), the distributions for both $K=0$ (c(e)) and $K=1$ (d(f)) are shown. 
The arrows below each figure indicate the position of the $\alpha$-cluster on the positive $Z$-axis.}
\label{fig:2n_distribution}
\end{figure}

To discuss the $2n$-cluster development in the LED states, we investigate the $2n$-cluster distribution in the $\3-b$ configuration. 
We calculate the squared overlap of the $0_1^+$ and $1_{1,2,3}^-$ states obtained by the GCM calculation with each basis wave function 
$\hat{P}_{MK}^{J}\hat{P}^{\pi}\Phi_{\3-b}(D,r,\theta)$ for which $D$ is chosen for each GCM states 
so as to give the maximum overlap.
In Fig.~\ref{fig:2n_distribution}, the $2n$ distribution around the $2\alpha$ 
in the $0_1^+$, $1_1^-$, $1_2^-$ and $1_3^-$ states are plotted on the $X - Z$ plane, 
where two $\alpha$-clusters are located at $(X,Y,Z)=(0,0,\pm D/2)$. 
The squared overlap with the $K=0$ and $K=1$ components are shown for the $0_1^+$ and $1^-_1$ states,
respectively, because they are dominant components.
For the $1_2^-$ and $1_3^-$ states, the results for the 
$K=0$ and $K=1$ components are shown because both components significantly contribute to these states.
 
In the TD state ($1_1^-$), the $2n$ concentrates in the region close to the $\alpha$ core and forms
a compact $\Hesix$ cluster (see Fig.~\ref{fig:2n_distribution} (b)).
It should be stressed that the TD state 
has the dominant $K=1$ component and is different from the  $K=0$ component
for the naive expectation of the $\alpha+\Hesix(0^+)$ cluster state in the relative $P$($L=1$) wave.
Instead, this state has a deformed $\Hesix$ cluster with a tilted orientation 
from the $\alpha - \alpha$ axis, and can be also understood by the single-particle excitation of the 
valence neutron as discussed in Refs.~\cite{Itagaki:1999vm,Ito:2003px}.

For the  $E1$ state ($1_2^-$) and the CD state ($1_3^-$) (Figs.~\ref{fig:2n_distribution} (c)-(f)), one can see the large $2n$-cluster development.
The $E1$ state ($1_2^-$) has the dominant
$K=0$ component with the remarkable $2n$ 
distribution in the region around $(X,Z)=(2 - 3~\textrm{fm},\ 3~\textrm{fm})$, (Fig.~\ref{fig:2n_distribution} (c)), whereas
the CD state ($1_3^-$) contains the $K=1$ component with the significant 
$2n$ distribution in the region around $(X,Z)=(4~\textrm{fm}, 1~\textrm{fm})$ (Fig.~\ref{fig:2n_distribution} (e)) much far from the $\alpha - \alpha$ axis.
It is found that the $E1$ state has the remarkably developed $\alpha - \alpha$ at 
$D=5$~fm and the $2n$-cluster with $K=0$
and the CD state has the moderately developed  $\alpha - \alpha$ at $D=4$~fm and the largely 
developed $2n$-cluster with $K=1$.
From these results, one can interpret, at the leading order, the $E1$ and CD states as 
relative $P$-wave excitations of the $\alpha  - (\alpha+2n)$ and $2n -  (2\alpha) $ cluster modes, respectively. 
However, the $E1$ and CD states also involve the other $K$ components (see Fig.~\ref{fig:2n_distribution} (d) and (f)). 
Because of the fragile nature of 3-body dynamics in $\3-b$,  
the $K$-mixing and also the $\2-b$ coupling occur in the final GCM states.

Let us discuss how the cluster developments in the $\3-b$ configurations contribute to the dipole transition strengths
of the $1_2^-$ and  $1_3^-$ states. 
As shown in the previous section, the $E1$ operator excites both states, 
whereas the CD operator strongly excites the $1_3^-$ state but not the $1_2^-$ state. 
The weak CD strength of the $1_2^-$ state seems to contradict the naive expectation that a developed cluster state could have the strong CD transition strength.
To clarify the properties of $E1$ and CD transitions of these states, 
we here focus on the typical two modes that dominantly contribute to the  $1_2^-$ and  $1_3^-$ states. 
One is the  $\alpha  -  (\alpha+2n)$ $K=0$ mode and the other is the $2n - (2\alpha)$ $K=1$ mode. 
The former is characterized by the large distance $D$ of the 
$\alpha$-cluster development and roughly understood by the 
$P$-wave excitation of the $\alpha - (\alpha+2n)$ relative motion (the $\alpha$-cluster development from the ($\alpha+2n$)).
The latter is the spatially extended $2n$-cluster distribution far from the $\alpha+\alpha$ and 
correspond to the $P$-wave excitation of the $2n - (2\alpha)$ relative motion (the $2n$-cluster development from the $2\alpha$).

For the CD transition, both of the two modes, $2n - (2\alpha)$ and $\alpha - (\alpha+2n)$, 
significantly contribute to the strength because of the developed cluster structures. 
The $1_2^-$ and  $1_3^-$ states are described by the linear combination of these two modes. 
In the  $1_3^-$ state, the $2n - (2\alpha)$ and $\alpha - (\alpha+2n)$ modes are superposed in phase
and coherently contribute to the CD transition strength, but in the  $1_2^-$ state they 
cancel the CD transition strength with each other because these configurations are superposed out of phase.
For the  $E1$ transition strength, 
the $2n - (2\alpha)$ mode gives larger contribution 
because of the large proton-number asymmetry between  $2n$ and $2\alpha$. 
However, $\alpha - (\alpha+2n)$ gives small contribution because of the 
smaller proton-number asymmetry and also the small 
overlap with the ground state.
The $2n - (2\alpha)$ component dominantly contributes to the $E1$ transition strengths of both the  $1_2^-$ and  $1_3^-$ states.
Quantitatively, the $1_3^-$ state has the largely developed $2n$-cluster component, with the smaller overlap with the ground state,
and therefore it has the relatively weaker $E1$ transition strength than the  $1_2^-$ state.

\begin{figure}[!h]
\begin{center}
\includegraphics[width=10cm]{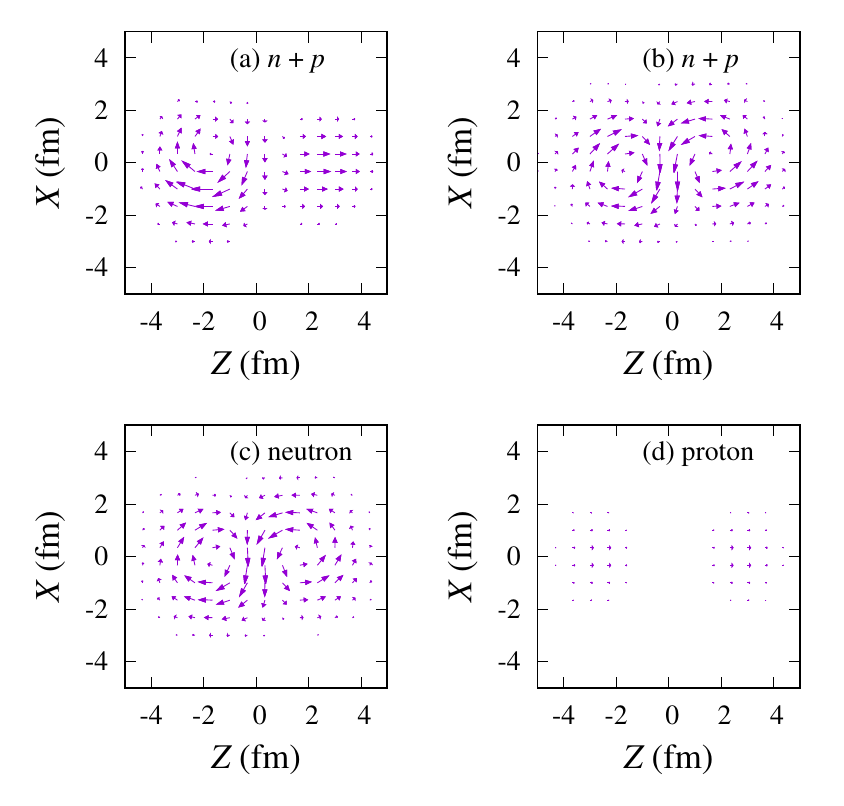} 	
\end{center}
\caption{(color online) Transition current densities $\delta \vector{j}(\vector{r})$ for the lowest dipole transitions
$0^+_1\to 1^-_1$. 
(a) $\delta \vector{j}(\vector{r})$ of the nuclear matter 
in the intrinsic state before the parity projection.
 $\delta \vector{j}(\vector{r})$ of the nuclear 
(b) matter, (c) proton, and (d) neutron densities parts
in the intrinsic state after the parity projection.
The vector multiplied by 400 is plotted on the $X-Z$ plane. 
}
\label{fig:toroidal}
\end{figure}

\subsection{Vortical nature of the TD state}\label{sec:toroidal}
The strong TD transition strength of the TD state ($1^-_1$) indicates the vortical nature of this state
because the TD operator can sensitively probe the nuclear vorticity as pointed out in Refs.~\cite{0954-3899-29-4-312,Kvasil:2011yk}.
We discuss the vortical nature in the transition current density of the $1^-_1$ state.

Strictly speaking, it is difficult to define the intrinsic state of the physical $1^-_1$ state obtained in the GCM calculation 
because the state is expressed by superposition of many configurations projected onto the parity and angular-momentum eigenstates.
We here consider the dominant configurations of the $0^+_1$ and $1_{1}^-$ states 
as approximate intrinsic states ($|0^+_{1,\textrm{int}}\rangle $ and $|1^-_{1,\textrm{int}}\rangle$)  and calculate the transition current density between 
them in the intrinsic frame. 
We take the $\2-b$ configuration at $D = 3$ fm for the  $0_1^+$ and that at $D = 4$ fm for the $1_{1}^-$ state. 
The two neutron configurations are taken to be  $(p_{x})^2$ and $(p_{x} + p_z)^2/2$ for the $0_1^+$ and $1_{1}^-$ states, respectively,
so that these configurations projected on the $J^\pi$ eigenstates 
have $70 - 90~\%$ overlap with the $0^+_1$ and $1_{1}^-$ states of the full GCM calculation.
The transition current density of the initial $|i\rangle$ and final $|f\rangle$ states is given as 
$\delta \vector{j}(\vector{r}) = \langle f | \vector{j}_{\textrm{nucl}}(\vector{r}) | i \rangle$, 
where $\vector{j}_{\textrm{nucl}}$ is the nuclear convection current density represented by \eqref{eq:current}.
We calculate $\delta \vector{j}(\vector{r})$ for 
$|i\rangle= |0^+_{1,\textrm{int}}\rangle$ and $|f\rangle=|1^-_{1,\textrm{int}}\rangle$
(before the parity projection) and also that for 
$|i\rangle=\hat{P}^+ |0^+_{1,\textrm{int}}\rangle$ and $|f\rangle=\hat{P}^-|1^-_{1,\textrm{int}}\rangle$
(after the parity projection).

The calculated transition current densities before and after the parity projection are plotted on the $X - Z$ plane in Fig.~\ref{fig:toroidal}. 
As clearly seen, the neutron transition current shows vortices
in the left and right sides on the $\alpha - \alpha$ ($X=Y=0$) axis. 
This is consistent with the result of our previous paper~\cite{Kanada-Enyo:2017uzz}.
In the $\2-b$ configuration, these vortices are generated by the $2n$ oscillation in the $^6$He cluster from $(p_x)^2$ into 
$(p_x+p_z)^2/2$ configurations, which corresponds to the tilting mode of the deformed $^6$He cluster.
In principle, we should not call this excitation mode in the TD state "toroidal dipole mode", but call it 
"vortical dipole (VD) mode" with a more general terminology because of following discussion.

\begin{figure*}[!h]
\begin{center}
\includegraphics[width=\hsize]{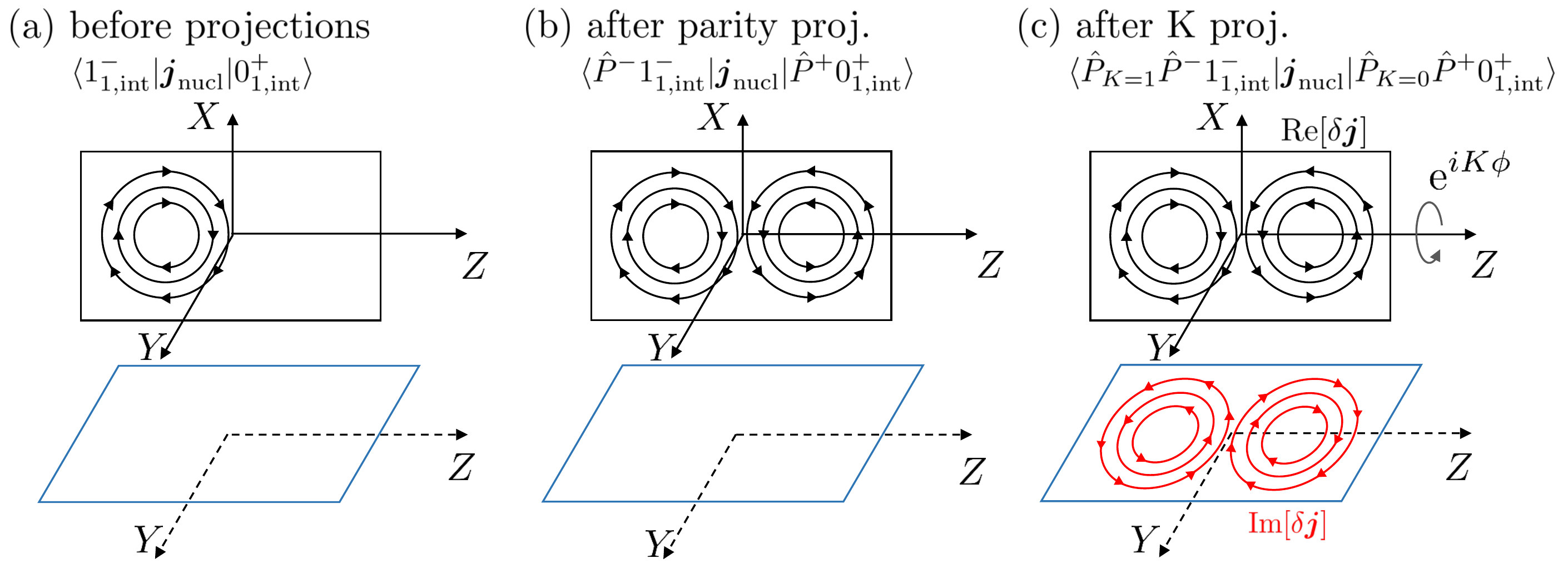} 	
\end{center}
\caption{(color online) Schematic figures of  the nuclear current in the VD mode. 
The sketch of the transition current density $\delta \vector{j}$ before the parity and $K$ projections, (b) that 
after the parity projection but before the $K$ projection, and (c) that after the parity and $K$ projections. 
The  real  part of $\delta \vector{j}$ on the $Z-X$ plane at $Y=0$ is shown in the upper panels, 
and the imaginary part of $\delta \vector{j}$ at $X=0$ on the $Z-Y$ plane is projected on the lower panels.
}
\label{fig:toroidal_schematic}
\end{figure*}

For an intuitive understanding, it is worth to describe the nuclear vorticity of the VD mode in the 
intrinsic states before and after the parity and $K$ projections. 
Schematic figures for the nuclear current before and after the parity and $K$ projections
are illustrated in Fig.~\ref{fig:toroidal_schematic}.
Let us start from the intrinsic states ($|0^+_{1,\textrm{int}}\rangle$ and $|1^-_{1,\textrm{int}}\rangle$) of the $\2-b$ cluster structure 
before the parity and $K$ projections (Fig.~\ref{fig:toroidal_schematic}~(a)), where the parity and axial symmetries are broken. 
In the tilting oscillation of the $^6$He cluster, two valence neutrons produce 
the surface neutron current around the $\alpha$ core and create a single vortex in left-side on the $\alpha - \alpha$ axis. 
In the parity-projected states ($\hat P^+|0^+_{1,\textrm{int}}\rangle$ and $\hat P^-|1^-_{1,\textrm{int}}\rangle$),
the parity symmetry is restored and the anti-vortex is generated in the right-side by the duplication of the vortex.  
Before the $K$ projection, the transition current density is zero on the $Z-Y$ plane at $X=0$ (the lower panel of Fig.~\ref{fig:toroidal_schematic}~(b)).
After the $K=1$ projection, the axial symmetry is restored in the $\hat P_{K=0} \hat P^+|0^+_{1,\textrm{int}}\rangle$ 
and $\hat P_{K=1} \hat P^-|1^-_{1,\textrm{int}}\rangle$ states with the phase factor $\exp(-i  K\phi)$.
Then, the nuclear current in the $0^+\to 1_1^-$ transition with $K=1$ has the periodic phase of $\exp(- i \phi)$ around the $\alpha - \alpha$ axis.
As a result, the imaginary part of the transition current appears in $Z-Y$ plane (see the lower panel of Fig.~\ref{fig:toroidal_schematic}~(c)).

It should be noted that the VD mode ($1^-_1$) of $\Beten$ is dominantly the $K=1$ dipole excitation on the prolate deformation, 
and its vortical current does not show 
the torus shape which has been originally suggested for the $K=0$ dipole excitation in spherical or axial symmetric nuclei. 
(It is mathematically obvious that the ideal torus shape can be seen only in the $K=0$ dipole excitation.)
Similar vortical mode is discussed in prolately deformed ${}^{24}$Mg in Ref.~\cite{Nesterenko:2017rcc}.
In this paper, we call the $1^-_1$ state TD state just because it is strongly excited by the TD operator because of the 
vortical nature in the dipole excitation. 
The $K=1$ VD mode is a new phenomena of nuclear vorticity peculiar to prolately deformed systems.

\section{Summary and outlook} \label{sec:summary}
We investigated the LED excitation modes in $\Beten$ based on the GCM with $\2-b$ and $\3-b$ cluster models.
In $E_{x} < 15$ MeV, we obtained three LED states.
The remarkable TD strength is obtained in the $1_1^-$ state regarded as the 
VD mode, 
in which tilting motion of the deformed $\Hesix$ cluster induces the toroidal nuclear current. 
The significant $E1$ strengths are obtained in the $1^-_2$ and $1^-_3$ states, while the
strong CD strength is obtained only in the $1_3^-$ state of the three LED.
The developed $\3-b$ cluster structures are found in the $1^-_2$ and $1^-_3$ states. 
In particular, two modes in the  $\3-b$ clustering dominantly contribute to these two LEDs.
One is the $P$-wave excitation of the $\alpha - (\alpha+2n)$ cluster mode, and the other is 
that of the $2n - (2\alpha)$ cluster mode. These two cluster modes describe the 
$E1$ and CD transition properties of the $1^-_2$ and $1^-_3$ states.
The remarkable $E1$ strengths in the low-energy region are predominantly contributed by the $2n - (2\alpha)$ cluster mode, which can not be described by the $\2-b$ cluster model. 

In comparison with experimental spectra, 
the VD mode is assigned to the experimental $1^-_1$ state at  5.96~MeV. 
The $1^-_2$ and $1^-_3$ states are theoretical predictions of the present calculation. 
Since the present calculation is a bound state approximation and omits decay widths of resonance states, 
the LED states were obtained as discrete levels. Even though the states should have decay widths, 
significant $E1$ and CD strengths can be expected in the low-energy region 
corresponding to the predicted LED modes in the present calculation.

In the present study, it was found that 
the cluster structure and valence neutrons play important roles
in the LED excitations in  $^{10}$Be.
The $2n$-cluster mode remarkably contributes the $E1$ and CD strengths.
Moreover, the TD state is produced by the tilting motion of the $2n$-cluster.
Also the $\alpha$-cluster mode was found to give the important contributions to the CD strengths. 
Similar phenomena in the LED excitations are expected in
other prolately deformed nuclei with valence neutrons such as $^{20}$O and $^{22}$Ne. 
The vortical LED mode in prolately deformed systems is one of the new interesting phenomena, 
and can be a key physics to clarify isospin properties of LED excitations in 
neutron-rich nuclei. 

\section*{Acknowledgment}

The author thanks to Dr.~Nesterenko, Dr.~Morita, and Dr.~Chiba for fruitful discussions.
The computational calculations of this work were performed by using the
supercomputer in the Yukawa Institute for theoretical physics, Kyoto University. 
This work was supported by 
JSPS KAKENHI Grant Nos.  18J20926, 26400270, 18K03617, and 16J05659.

\bibliographystyle{ptephy}
\bibliography{reference_10Be} 

\end{document}